\documentclass[showpacs,preprintnumbers,superscriptaddress,pre]{revtex4-1}
\usepackage{epsfig}
\usepackage{graphicx}
\usepackage{amsmath,amsfonts,amssymb}
\usepackage{pstricks}

\renewcommand{\epsilon}{\varepsilon}
\def\be{\begin{equation}}
\def\ee{\end{equation}}
\newcommand\bea{\begin{eqnarray}}
\newcommand\eea{\end{eqnarray}}

\def\arrowsite{\begin{picture}(45,5)(-2,-2)
\put(0,0){\circle*{5}}
\put(2.5,0){\vector(1,0){10}}
\put(12.5,0){\line(1,0){15}}
\put(37.5,0){\vector(-1,0){10}}
\put(40,0){\circle*{5}}
\end{picture}}

\def\arrowroot{\begin{picture}(45,5)(-2,-2)
\put(0,0){\circle*{5}}
\put(2.5,0){\vector(1,0){10}}
\put(15,0){\circle{5}}
\put(25,0){\circle{5}}
\put(37.5,0){\vector(-1,0){10}}
\put(40,0){\circle*{5}}
\end{picture}}

\begin{document}

\title{Exact finite-size corrections and corner free energies for the $c=-2$ universality class}

\author{Nickolay Izmailian}
\email{izmail@yerphi.am; ab5223@coventry.ac.uk}
\affiliation{Applied Mathematics Research Center, Coventry University, Coventry CV1 5FB, UK}
\affiliation{Yerevan Physics Institute, Alikhanian Brothers 2, 375036 Yerevan, Armenia}

\author{Ralph Kenna}
\email{r.kenna@coventry.ac.uk}
\affiliation{Applied Mathematics Research Center, Coventry University, Coventry CV1 5FB, UK}

\author{Wenan Guo}
\email{waguo@bnu.edu.cn}
\affiliation{Department of Physics, Beijing Normal University, Beijing 100875, China}

\author{Xintian Wu}
\email{wuxt@bnu.edu.cn}
\affiliation{Department of Physics, Beijing Normal University, Beijing 100875, China}

\date{\today}

\begin{abstract}
We consider
(a) the partition functions of the anisotropic dimer model on the rectangular $(2M-1) \times (2N-1)$ lattice with free and cylindrical boundary conditions with a single monomer residing on the boundary and
(b) the partition function of the anisotropic spanning tree on an $M \times N$ rectangular lattice with free boundary conditions.
We express (a) and (b) in terms of a principal partition function with twisted boundary conditions.
Based on these expressions, we derive the exact asymptotic expansions of the free energy for both cases (a) and (b).
We confirm the conformal field theory prediction for the corner free energy of these  models, and find the central charge  is $c = - 2$.
We also show that the dimer model on the cylinder with an odd number of sites on the perimeter exhibits the same finite-size corrections as on the plane.
\end{abstract}


\maketitle

\section{Introduction}
\label{Introduction}

Finite-size scaling has been of interest to scientists working
on a variety of critical systems, including spin models, percolation models, lattice gauge models, spin
glass, etc. \cite{privman1990}. The properties of the associated corrections to theoretical predictions for the behavior of idealized infinite systems play increasingly important roles in improving understanding  of real statistical systems in the critical regime. Therefore, in recent decades there have many
investigations on finite-size scaling, finite-size corrections, and boundary effects for model systems.

To fully understand such effects, analyses which can be carried out without delivering numerical errors are of particular importance. These include systems such as the Ising model \cite{ferdinand1969,ising,izm2002}, spanning-tree models \cite{tseng2003}, dimer models \cite{tseng2003,dimer,izm2003,izm2005,izm2007c,izm2013}, the critical dense polymer model \cite{pearce2010,izm2013}, resistor network \cite{izm2010}, Hamiltonian walk \cite{dupl1988}, critical Potts model \cite{hu1999} etc., which allow for exact studies.

Models with exact solutions, therefore, have key roles in understanding the forms of
finite-size scaling. Ferdinand and Fisher stimulated such studies \cite{ferdinand1969} by performing a finite-size analysis of Onsager's exact solution \cite{onsager1944} of the two-dimensional Ising model on finite-size rectangular lattices with periodic boundary conditions.
Although the  solution for the Ising model  with free boundaries  is still lacking, exact solutions for a variety of different models with various boundaries have been obtained and studied intensively.

Many critical systems have been shown to have local scale invariance, so that their scaling limits can be described by conformal field theory. Such a theory is parameterized by the value of its central charge $c$, which itself is related to the finite-size corrections to the critical free energy.
For critical two-dimensional systems, it has long been known  that the free energy contains a term of order $O(\ln{L})$ due to corner singularities, where $L$ is the typical size of the system, with a universal prefactor proportional to the central charge \cite{cardy1988}.
The study of statistical systems in the presence of such corner singularities has emerged as a topic in its own right -- one which is increasingly gained in importance  \cite{wu2012,wu2013,wu2014,corner,cardy1988,kleban1991}.

Several years ago an efficient bond-propagation algorithm was
developed for computing the partition function of the Ising model
with free edges and corners in two dimensions  \cite{loh2007}.
With this algorithm,  calculations have been carried out on various lattices
and the results are  accurate to a remarkable margin of $10^{-26}$ \cite{wu2012,wu2013,wu2014}.
Fitting the standard finite-size scaling formulae to associated data allowed the edge and corner terms for the free energy to be obtained very accurately.
For example, from the corner term for the rectangular lattice (comprising rectangular elementary plaquette) \cite{wu2012}
and from the corner terms for triangular, rhomboid, trapezoid, hexagonal and rectangular lattices (each comprising elementary triangular plaquette) \cite{wu2013}, the central charge of the Ising model was estimated to be $c=0.5\pm 1\times 10^{-10}$, compared with the conformal-field-theory result $c=0.5$ \cite{cardy1988}.

Conformal invariance implies that on a finite   lattice with free boundaries the critical free energy has the generic form \cite{kleban1991}
\begin{equation}
F={\cal{S}} f_{\rm{bulk}}+ {\cal{P}} f_{\rm{surf}}+ f_0 + O(1/{\cal{S}}) \label{F}.
\end{equation}
Here ${\cal{S}}$ represents the area of the lattice and $f_{\rm{bulk}}$ is the free energy per unit area.
The second term represents contributions from the lattice perimeter ${\cal{P}}$, with
$f_{\rm{surf}}$ the associated free energy per unit edge length.

In general, the coefficients $f_{\rm{bulk}}$ and $f_{\rm{surf}}$ are non-universal, but the
coefficient $f_0$ is supposed \cite{privman1984} to be universal, depending only on the
shape of the system and, possibly, the nature of the boundary conditions. In some
two-dimensional geometries, the value of $f_0$ is known \cite{blote1986} to be simply related to the
conformal anomaly number c of the theory. Cardy and Peschel \cite{cardy1988} have shown that corners on the boundary induce a trace anomaly in the stress tensor. This gives rise to a term in $f_0$ proportional to $\ln \cal{S}$, where $\cal{S}$ is the area of the domain. Later, Kleban and Vassileva \cite{kleban1991} have shown that in rectangular geometry in addition to corner contribution proportional to $\ln \cal{S}$ the term $f_0$ contains a term depending on the aspect ratio, namely, the term $f_0$ contains the universal part $f_{\rm{univ}}$ given by
\begin{equation}
f_{\rm{univ}}=-\frac{c}{8}\ln{\cal{S}}+\frac{c}{4}\ln\left[\eta(q)\eta(q')\right] . \label{funiv}
\end{equation}
Here $c$ is the central charge, $q=\exp{(-2\pi\xi)}$, $q'=\exp{(-2\pi/\xi)}$, $\xi$ is the aspect ratio and $\eta$ the Dedekind eta function.

In this paper we  derive exact asymptotic expansions, to arbitrary order, for  the free energies of  critical systems described by logarithmic conformal field theory  with central charge $c = -2$.
Such systems are realized as the dimer model on a rectangular lattice, the Abelian sandpile model, the spanning tree, Hamiltonian walks on a Manhattan lattice, rational triplet theory, symplectic fermions, the traveling salesman problem, as well as branching polymers.
The calculation of the central charge, based on  finite-size corrections for the dimer model on the rectangular lattice, has led to some confusion in the literature, due to the (mis)interpretation of finite-size corrections in terms of the central charge rather than the effective central charge \cite{tseng2003,chakr2002}. Only quite recently it has been shown \cite{izm2005,izm2007c,izm2013} that the central charge for the dimer model is $c= - 2$.
In particular, we consider (a) the anisotropic dimer model on a $(2M-1) \times (2N-1)$ rectangular lattice with free and cylindrical boundary conditions with a single
monomer on the boundary and  (b) the anisotropic spanning tree on the $M \times N$ rectangular lattice with free boundary conditions. We show that the exact asymptotic expansion for the free energy for these models can be written as
\begin{equation}
F={\cal{S}} f_{\rm{bulk}}+2 {\cal N}
f_{1s}(x,y)+2 {\cal M} f_{2s}(x,y) + f_0(z \xi)+
\sum_{p=1}^\infty \frac{f_{p}(z \xi)}{{\cal{S}}^{p}},
\label{Fnew}
\end{equation}
where  ${\cal{S}} = {\cal M} {\cal N}$,
$f_{1s}$ and $f_{2s}$ are the free energies per unit edge length in the horizontal  and vertical directions  respectively, along which $x$ and $y$ are the dimer weights  with $z = x/y$. The quantities $\cal{M}$ and $\cal{N}$ are functions of the physical dimensions of the lattice and the aspect ratio is $\xi = {\cal N}/{\cal M}$.
{\sl All} coefficients in the expansion (\ref{Fnew}) are expressed through analytical functions.
The correspondences  between ${\cal M}$ and ${\cal N}$ in Eq.(\ref{Fnew}) and the lattice dimensions for system (a) and (b) are summarized as:
$$
({\cal M}, {\cal N})=\left\{\begin{array}{lcl}
(2 M, 2 N) \hspace{1cm} \mbox{ for the dimer model with free
boundary conditions}, \label{MNfree}\\
 (2 M -1, 2 N) \hspace{0.5cm} \mbox {for the dimer model on cylinder,} \label{MNcyl}\\
 (M, N) \hspace{1.5cm} \mbox {for the spanning tree.} \label{MNsp}
\end{array}
\right.
$$

We show that $f_0$ contains the universal part $f_{\rm{univ}}$ given by Eq. (\ref{funiv}).
This confirms the conformal field theory prediction for the corner free energy in
models for which the central charge is $c = - 2$.


The paper is organized as follows. In Sec. II we show how to express
(a) the partition function  for the dimer model on a $(2M-1) \times (2N-1)$ rectangular lattice with free and cylindrical boundary conditions with a single monomer on the boundary and (b) the partition function for the spanning
tree on an $M \times N$ rectangular lattice with free boundary
conditions in the form of a partition function with twisted boundary
conditions. In Sec. III asymptotic expansions of the free energies are presented.
The results are summarized and discussed in Sec. IV.

\section{Partition function of the dimer and spanning-tree models}
\label{PartitionDimerSpanning}

Consider a anisotropic dimer model on a finite rectangular lattice with an odd number of rows and an odd number of columns.
The lattice is planar if there are free boundary conditions in both directions. It is cylindrical if there are periodic boundary conditions in the horizontal direction, for example, and free boundary conditions in the vertical direction.
The partition function for the anisotropic dimer model is given by
\begin{equation}
Z(x,y)=\sum x^{n_h} y^{n_v},
\label{ZDimer}
\end{equation}
where the summation is taken over all dimer covering configurations.
Here, $n_h$ and $n_v$ are the number of horizontal and vertical dimers whose weights are  $x$ and $y$ respectively.
It has been shown that the exact partition functions of the anisotropic dimer model on
finite rectangular lattices with free, cylindrical, toroidal, M\"{o}bius-strip
and Klein-bottle boundary conditions can be expressed in terms of
the principal object
\begin{equation}
Z^2_{\alpha,\beta}(z,M,N)=
\prod_{n=0}^{{N}-1}\prod_{m=0}^{M-1}4\left[\textstyle{ \;z^2
\sin^2{\frac{(m+\alpha) \pi}{{M}}}+
\sin^2{\frac{(n+\beta)\pi}{N}}}\right], \label{Zab}
\end{equation}
where $(\alpha, \beta) = (1/2, 0),~(0, 1/2)$ or $(1/2, 1/2)$ and $z=x/y$.
Here, $M$ and $N$ are related to the lattice dimensions, the precise details depending on the lattice geometry in question \cite{izm2003}.

It is clear that $Z_{0,0}(z,M,N)$ vanishes due to the zero mode at $(m,n)=(0,0)$.
In what follows, therefore, we remove the zero mode, and when  $\alpha = \beta=0$ replace $Z_{0,0}(z,M,N)$ in Eq.(\ref{Zab}) by
\begin{equation}
Z^2_{0,0}(z,M,N)=
\prod_{n=0}^{{N}-1}{\prod_{m=0}^{M-1}{\!\!\!\!}^{~^\prime}}4\left[\textstyle{
\;z^2 \sin^2\frac{m \pi}{{M}}+ \sin^2\frac{n \pi}{N}}\right],
\label{Z00}
\end{equation}
where the prime on the product denotes the restriction $(m,n) \ne
(0,0)$.

The general theory for the asymptotic expansion of $Z_{\alpha,\beta}(z,M,N)$ for $(\alpha,\beta) \ne (0,0)$ appearing in the anisotropic dimer model has been given in \cite{izm2003}. In this paper we will present the asymptotic expansion of $Z_{0,0}(z,M,N)$.

In this section, we consider (a) the anisotropic dimer model on $(2M-1) \times
(2N-1)$ rectangular lattices with free and cylindrical boundary conditions with
a single monomer on the boundary and (b) the anisotropic spanning tree on an $M \times N$ lattice with free boundary conditions. The aim of the section is to show that the partition functions for both (a) and (b) can be written in terms of the principal mathematical object appearing in Eq.(\ref{Z00}).

\subsection{Partition function of the dimer models}
\label{PartitionDimer}

We consider, in turn, two sets of boundary conditions for the dimer model on $(2M-1) \times
(2N-1)$ rectangular lattices.

\subsubsection{Dimer model on the rectangular lattice with free boundary condition}
\label{dimerplane1}

The exact partition function for the dimer model on a $(2M-1) \times
(2N-1)$ rectangular lattice with free boundary conditions and with a
single monomer on the boundary is given by \cite{tseng2003,Wu}
\begin{eqnarray}
Z_{2M-1,2N-1}^{\rm{free}}&=&x^{M-1}y^{N-1}\prod_{n=1}^{N-1}\prod_{m=1}^{M-1}
4\left[\textstyle{ \;x^2\cos^2\frac{m \pi}{2 M}+ y^2 \cos^2\frac{n
\pi }{2 N} \;}\right] \nonumber\\
&=&z^{M-1}y^{2 M N-M-N}\prod_{n=1}^{N-1}\prod_{m=1}^{M-1}
4\left[\textstyle{ \;z^2\cos^2\frac{m \pi}{2 M}+  \cos^2\frac{n
\pi }{2 N} \;}\right]. \label{Zfree}
\end{eqnarray}
We change the variables $n \to N-n$ and $m \to M-m$ to write the partition function
as
\begin{equation}
Z_{2M-1,2N-1}^{\rm{free}} =
z^{M-1}y^{2 M N-M-N}\prod_{n=1}^{N-1}\prod_{m=1}^{M-1}
4\left[\textstyle{ \;z^2\sin^2\frac{m \pi}{2 M}+  \sin^2\frac{n
\pi }{2 N} \;}\right]. \label{Zfree1}
\end{equation}
It is easy to show that $f(2N-n,m)=f(n,2M-m)=f(n,m)$, where
\begin{equation}
f(n,m)=4\left[\textstyle{ \;z^2\sin^2\frac{m \pi}{2 M}+
\sin^2\frac{n \pi}{2 N} \;}\right].
\label{transf1}
\end{equation}
This allows us to express the double product
$\prod_{n=0}^{2N-1}{{\prod^{\prime}}_{m=0}^{M-1}} f(n,m)$
in terms of the simpler
$\prod_{n=1}^{N-1} \prod_{m=1}^{M-1}f(n,m)$ through
\begin{equation}
\prod_{n=0}^{2N-1}
{\prod_{m=0}^{M-1}{\!\!\!\!}^{{~^\prime}}}
f(n,m)=\frac{\prod_{n=1}^{2N-1}f(n,M) f(n,0)
\prod_{m=1}^{2M-1}f(N,m)f(0,m)}{f(N,M)}
\left[\prod_{n=1}^{N-1}
\prod_{m=1}^{M-1}f(n,m)\right]^4, \label{transf2}
\end{equation}
with $f(N,M)=4(1+z^2)$.
With the help of the identities \cite{GradshteinRyzhik}
\begin{equation}
4 \sinh^2\left(M\; \omega \right) = \prod_{m=0}^{M-1}4\textstyle{
\left[~\!\sinh^2\omega + \sin^2\frac{m \pi }{M}\right]},
\label{identity1}
\end{equation}
and
\begin{equation}
\prod_{m=1}^{M-1}4 \sin^2\frac{m \pi}{M}=M^2, \label{identity2}
\end{equation}
the products $\prod_{n=1}^{2N-1}f(n,M)f(n,0)$ and
$\prod_{m=1}^{2M-1}f(N,m)f(0,m)$ can be written as
\begin{equation}
\prod_{n=1}^{2N-1}f(n,M)f(n,0)=\frac{4 N^2}{z^2} \sinh^2{\left(2
N{\rm arcsinh}\, z \right)},
\label{prod1}
\end{equation}
and
\begin{equation}
\prod_{m=1}^{2M-1}f(N,m)f(0,m) = 4 M^2 z^{8 M  - 2}
\sinh^2{\left(2 M{\rm arcsinh}\, \frac{1}{z}\right)},
\label{prod2}
\end{equation}
respectively.
Now, using Eqs.(\ref{Z00}), (\ref{Zfree1})-(\ref{transf2}),
(\ref{prod1}) and (\ref{prod2}) the partition function for dimers
with free boundary conditions can finally  be written as
\begin{eqnarray}
Z_{2M-1,2N-1}^{\rm{free}} =Q \; Z^{1/2}_{0,0}(z,2M,2N),
\label{Zfree2}
\end{eqnarray}
in which
\begin{eqnarray}
Q = \frac{y^{2M N-M-N}}{z^M}\frac{(1+z^2)^{1/4}}{\sqrt{2 M N
\sinh{\left(2 N{\rm arcsinh}\, z \right)}\sinh{\left(2 M{\rm
arcsinh}\, 1/z \right)}}}.
\label{Zfree21}
\end{eqnarray}

\subsubsection{Dimer model on the rectangular lattice with cylindrical boundary conditions}
\label{dimercyl1}

The exact partition function for the dimer model on a $(2M-1) \times (2N-1)$ rectangular lattice, with cylindrical boundary conditions, with a single monomer on the boundary  is given by \cite{wu2011}
\begin{eqnarray}
Z_{2M-1,2N-1}^{\rm{cyl}}&=&x^{M-1}y^{N-1}\prod_{n=1}^{N-1}\prod_{m=1}^{M-1}
4\left[\textstyle{ \;x^2\sin^2\frac{ 2 m \pi}{2 M -1}+ y^2
\cos^2\frac{n
\pi }{2 N} \;}\right]
\nonumber\\
&=&z^{M-1}y^{2 M N-M-N}\prod_{n=1}^{N-1}\prod_{m=1}^{M-1}
4\left[\textstyle{ \;z^2\sin^2\frac{2 m \pi}{2 M - 1}+
\cos^2\frac{n \pi }{2 N} \;}\right]. \label{Zcyl}
\end{eqnarray}
Using the transformation $n \to 2N-n$ and the relation
\begin{eqnarray}
\prod_{m=1}^{M-1} \left(a+\sin^2\frac{2 m \pi}{2M-1}\right)
=\prod_{m=1}^{M-1} \left(a+\sin^2\frac{m \pi}{2M-1}\right),
\label{relationodd1}
\end{eqnarray}
the partition function given by Eq. (\ref{Zcyl}) can be
expressed in the  form
\begin{equation}
Z_{2M-1,2N-1}^{\rm{cyl}} =
z^{M-1}y^{2M N-M-N}
\prod_{n=1}^{N-1}\prod_{m=1}^{M-1} 4\left[\textstyle{
\;z^2\sin^2\frac{m \pi}{2 M-1}+  \sin^2\frac{n \pi }{2 N}
\;}\right]. \label{Zcyl1}
\end{equation}
Following the same procedure as in the case of free boundary
conditions, we  obtain
\begin{eqnarray}
Z_{2M-1,2N-1}^{\rm{cyl}} =R \; Z^{1/2}_{0,0}(z,2M-1,2N), \label{Zcyl2}
\end{eqnarray}
with
\begin{eqnarray}
R = \frac{y^{2M N-M-N}}{z^{M-1/2} \sqrt{2 N (2M-1) \sinh{\left[(2
M-1){\rm arcsinh}\, 1/z \right]}}}. \label{Zcyl21}
\end{eqnarray}

\subsection{Partition function of the spanning tree model}
\label{PartitionSpanning}

Let us now consider the problem of enumerating weighted spanning trees on the $M \times N$ rectangular lattice with free boundary conditions.
The problem of enumerating spanning trees on a graph was first considered by Kirchhoff in his analysis of electrical networks \cite{Kirchhoff}.
The enumeration of spanning trees involves the evaluation of the tree generating function
\begin{equation}
Z_{M,N}^{\rm{span}}=\sum_T u^{n_h}v^{n_v},
\label{span}
\end{equation}
where weights $u$ and $v$, respectively, are assigned to the edges in the horizontal and vertical direction. The summation is taken over all spanning tree configurations $T$ on the lattice and, $n_h$ and $n_v$ are the numbers of edges in the spanning tree in the respective directions. For a rectangular $M \times N$ lattice with free boundaries, the exact partition function for the weighted spanning trees $Z_{M,N}^{\rm{span}}$  is given by \cite{tseng2000}
\begin{eqnarray}
Z_{M,N}^{\rm{span}}&=&\frac{1}{M
N}\prod_{m=0}^{M-1}\prod_{n=0 \above0pt  (m,n)\neq(0,0)}^{N-1}
2\left[\textstyle{ \;u\left(1-\cos\frac{m \pi}{M}\right)+ v\left(1- \cos\frac{n
\pi }{N}\right) \;}\right] . \label{Zsptr}
\end{eqnarray}
This partition function can be rewritten as
\begin{equation}
Z_{M,N}^{\rm{span}}=\frac{1}{M
N}\prod_{m=0}^{M-1}\prod_{n=0 \above0pt  (m,n)\neq(0,0)}^{N-1}
4\left[\textstyle{ \;u\sin^2\frac{m \pi}{2M}+ v\sin^2\frac{n
\pi }{2N} \;}\right]. \label{Zsptrtransf}
\end{equation}
Denoting $u=x/y$ and $v=y/x$ we arrive at
\begin{equation}
Z_{M,N}^{\rm{span}}=
\frac{1}{M N z^{M N-1}}\prod_{m=0}^{M-1}\prod_{n=0 \above0pt  (m,n)\neq(0,0)}^{N-1}
4\left[\textstyle{ \;z^2\sin^2\frac{m \pi}{2M}+\sin^2\frac{n \pi
}{2N} \;}\right] ,  \label{Zsptr2}
\end{equation}
where $z=x/y$.
$Z_{M,N}^{\rm{span}}$ can then be transformed to
\begin{eqnarray}
Z_{M,N}^{\rm{span}}&=&\frac{1}{M N z^{M N-1}}\left(\prod_{n=1}^{N-1}4\sin^2\frac{n \pi}{2N}\right)\left(\prod_{m=1}^{M-1}4z^2\sin^2\frac{m \pi}{2M}\right)\prod_{m=1}^{M-1}\prod_{n=1}^{N-1}
4\left[\textstyle{ \;z^2\sin^2\frac{m \pi}{2M}+\sin^2\frac{n \pi
}{2N} \;}\right] \label{Zsptr3}\\
&=&z^{-M N+2M-1}\prod_{m=1}^{M-1}\prod_{n=1}^{N-1}
4\left[\textstyle{ \;z^2\sin^2\frac{m \pi}{2M}+\sin^2\frac{n \pi
}{2N} \;}\right].\label{Zsptr4}
\end{eqnarray}
Here we again used the  identity (\ref{identity2}).
Compare the expression for the partition function of the spanning tree on the $M \times N$ lattice given by Eq. (\ref{Zsptr4}) with that of the dimer model on $(2M-1) \times (2N-1)$ lattice with free boundary conditions and single monomer on the boundary given by Eq. (\ref{Zfree1}).
This comparison delivers the identity
\begin{equation}
Z_{2M-1,2N-1}^{\rm{free}}=x^{M N-M} y^{M N-N} Z_{M,N}^{\rm{span}}, \label{DST}
\end{equation}
which proves a proposition  due to Tzeng and Wu   (Eq.(3.1.2) of Ref.\cite{tseng2003}).

Therefore, as in the case of the dimer model, the partition function of the spanning-tree model on an $M \times N$ rectangular lattice with free boundary conditions can be expressed in terms of $Z_{0,0}(z,M,N)$ as
\begin{eqnarray}
Z_{M,N}^{\rm{span}}&=&Q_1 Z_{0,0}^{1/2}(z,2M,2N), \label{Zsptr5}
\end{eqnarray}
where
\begin{equation}
Q_1
=
\frac{Q}{x^{M N-M}y^{M N-N}}
=\frac{1}{z^{M N}}\frac{(1+z^2)^{1/4}}{\sqrt{2 M N
\sinh{\left[2 N{\rm arcsinh}\, z \right]}\sinh{\left[2 M{\rm
arcsinh}\, 1/z \right]}}}. \label{Q11}
\end{equation}

Eqs.(\ref{Zfree2}), (\ref{Zcyl2}) and (\ref{Zsptr5}) give how the partition functions of (a) the dimer model on a $(2M-1)\times(2N-1)$ rectangular lattice with free and cylindrical boundary conditions, with a single monomer on the boundary,
and (b) of the spanning tree model on an $M \times N$
rectangular lattice with free boundary conditions
can be expressed in
terms of the principal object $Z_{0, 0}(z,M,N)$.
Based on such results, one can use the method proposed
by Ivashkevich, Izmailian, and Hu \cite{izm2002} to derive the
asymptotic expansion of the $Z_{0, 0}(z,M,N)$ in terms of
Kronecke'r double series \cite{Weil}, which are directly related to
elliptic $\theta$ functions (see Appendix \ref{Expansion Z(0,0)(z,M,N)}).

\section{Asymptotic expansion of free energy}
\label{free energy}
In Sec. II, we have shown that the partition functions of
the spanning tree and the dimer model  with free and cylindrical boundary conditions with a single monomer on the boundary can be expressed in terms of the principal partition function with twisted
boundary conditions $Z_{0, 0}(z,M,N)$ (see Eqs. (\ref{Zfree2}), (\ref{Zcyl2}) and (\ref{Zsptr5})). The
asymptotic expansion of the $Z_{0, 0}(z,M,N)$ is given in Appendix \ref{Expansion Z(0,0)(z,M,N)}.

After reaching this point, one can use Eq. (\ref{ExpansZab}) to write down all the terms
of the exact asymptotic expansions of the free energy, $F = - \ln Z$ for all models under consideration in the form of  Eq.(\ref{Fnew}).



\subsection{Dimer model on the rectangular lattice with  free and cylindrical boundary conditions}
\label{dimerplane}

The bulk free energy $f_{\rm{bulk}}$ for the dimer model on a $(2M-1) \times (2N-1)$ rectangular lattice for both free and cylindrical boundary conditions is given by
\begin{eqnarray}
f_{\rm{bulk}}&=&-\frac{1}{2}\ln{y}-\frac{1}{2\pi}\int_0^\pi\omega_z(x)dx
\nonumber\\
&=&-\frac{1}{2}\ln{y}-\frac{1}{2\pi}\int_0^\pi{\rm
arcsinh}{\left(z \sin{x}\right)}dx
\label{fbulk}\\
&=&-\frac{1}{2}\ln{y}-\frac{\Phi(-z^2,2,\frac{1}{2})}{4\pi},
\nonumber
\end{eqnarray}
where $\omega_z(x)$ is lattice dispersion relation defined in Eq. (\ref{SpectralFunction}) and $\Phi(x,s,\alpha)$ is the Lerch transcendent defined
in Eq.(\ref{phi}).
In particular for the isotropic dimer model ($z=1$), $\Phi(-1,2,{1}/{2})=4 G$, where $G$ is the
Catalan constant given in Eq.(\ref{Catelan}) as
$G=0.915\, 965\, 594 \dots$.
The surface free energy $f_{1s}$ and $f_{2s}$ defined by Eq.
(\ref{Fnew}) are
\begin{eqnarray}
f_{1s}^{\rm{free}}&=&\frac{1}{4}\ln{y}+\frac{1}{4}\ln(z+\sqrt{1+z^2}),
\label{f1surf}\\
f_{1s}^{\rm{cyl}}&=&0,
\label{f1scyl}\\
f_{2s}^{\rm{free}}&=&f_{2s}^{\rm{cyl}}=\frac{1}{4}\ln{y}+\frac{1}{4}\ln(1+\sqrt{1+z^2}).
\label{f2surf}
\end{eqnarray}
For the leading correction terms $f_0(z\xi)$ we obtain
\begin{eqnarray}
f_0^{\rm{free}}(z \xi)&=&\frac{1}{4}\ln{S}-\frac{1}{4}\ln{\xi}-\ln {\eta(i z\xi)}-\frac{3}{2}\ln{2}-\frac{1}{4}\ln{(1+z^2)}
\nonumber\\
&=&\frac{1}{4}\ln{S}-\frac{1}{2}\ln {\eta(i z\xi)\eta(i/(z\xi))}-\frac{3}{2}\ln{2}-\frac{1}{4}\ln{(1+z^2)}+\frac{1}{4}\ln z,
\label{f0free}
\end{eqnarray}
in which $\xi = \frac{N}{M}$,
\begin{eqnarray}
f_0^{\rm{cyl}}(z \xi)&=&\frac{1}{4}\ln{S}-\frac{1}{4}\ln{\xi}-\ln {\eta(i z\xi)}-\frac{1}{2}\ln{2}+\frac{1}{2}\ln{y} \nonumber\\
&=&\frac{1}{4}\ln{S}-\frac{1}{2}\ln {\eta(i z\xi)\eta(i/(z\xi))}-\frac{1}{2}\ln{2}+\frac{1}{2}\ln{y}+\frac{1}{4}\ln z,
\label{f0cyl}
\end{eqnarray}
in which $\xi = \frac{2N}{2M-1}$.
Here $\eta$ is the Dedekind  eta function and we use the behavior of the this function under the Jacobi transformation $\tau \to \tau'=-1/\tau$,
\begin{equation}
\eta(\tau')=\sqrt{-i \tau} \eta(\tau),
 \label{jacobi}
\end{equation}
for $\tau = i z \xi$.

For the subleading correction terms $f_p(z \xi)$ for $p=1, 2, 3,
..., $ we obtain
\begin{eqnarray}
f_p(z \xi)&=&\pi^{2p+1} \xi^{p+1}\frac{\Lambda_{2p}}{(2p)!}
\frac{ K_{2p+2}^{0,0}(i z \xi)}{2p+2}, \nonumber
\end{eqnarray}
where for free boundary conditions, we again use $\xi = \frac{N}{M}$ and for cylindrical boundary conditions $\xi=\frac{2N}{2M-1}$.
As an example, we list first few expansion coefficients $f_p(z \xi)$ for $p=1,~2$
\begin{eqnarray}
f_1(z\xi)&=& z(1+z^2)\frac{\pi^3 \xi^2 }{1080}\;(\theta
_2^4\theta _3^4+\theta _3^4\theta _4^4-\theta _2^4\theta _4^4),
\nonumber\\
f_2(z\xi)&=&\frac{\pi^5 \xi^3
}{12096}\left[\frac{z(1+z^2)(1+9z^2)}{5}+\frac{z^2(1+z^2)^2}{3}\frac{\partial}{\partial z}\right]
(\theta_2^4+\theta_3^4)(\theta_2^4-\theta_4^4)(\theta_3^4+\theta_4^4),
\end{eqnarray}
where  $\theta_i = \theta_i(z \xi)$ with $i=2,3,4$.

Note, that with the help of the identities
$$
\frac{\partial}{\partial \xi}\ln{{\theta}_3} =
\frac{\pi}{4}{\theta}_4^4+\frac{\partial}{\partial
\xi}\ln{{\theta}_2}, \qquad \mbox{and} \qquad
\frac{\partial}{\partial \xi}\ln{{\theta}_4} =
\frac{\pi}{4}{\theta}_3^4+\frac{\partial}{\partial
\xi}\ln{{\theta}_2}:
$$
$$
\frac{\partial}{\partial \xi}\ln{\theta_2} =
-\frac{1}{2}{\theta}_3^2 E, \qquad \mbox{and} \qquad
\frac{\partial E}{\partial \xi}
=\frac{\pi^2}{4}{\theta}_3^2{\theta}_4^4-\frac{\pi}{2}{\theta}_4^4 E,
$$
one can express all derivatives of the elliptic functions in terms
of the elliptic functions ${\theta}_2, {\theta}_3, {\theta}_4$ and
the elliptic integral of the second kind $E$.

\subsection{Spanning tree on the rectangular lattice with free boundary conditions}
\label{spannqplane}

The bulk free energy $f_{\rm{bulk}}^{\rm{span}}$ for the spanning tree on an $M \times N$ lattics with free boundaries is given by
\begin{eqnarray}
f_{\rm{bulk}}^{\rm{span}}&=&\ln{z}-\frac{2}{\pi}\int_0^\pi\omega_z(x)dx
\label{fbulk}\\
&=&\ln{z}-\frac{\Phi(-z^2,2,\frac{1}{2})}{\pi},
\nonumber
\end{eqnarray}
where $\Phi(x,s,\alpha)$ is the Lerch transcendent. The surface free energy for the spanning tree $f_{1s}^{\rm{span}}$ and $f_{2s}^{\rm{span}}$ in Eq. (\ref{Fnew}) are
\begin{eqnarray}
f_{1s}^{\rm{span}}&=&\frac{1}{2}\ln(z+\sqrt{1+z^2}),
\label{f1surfsp}\\
f_{2s}^{\rm{span}}&=&-\frac{1}{2}\ln{z}+\frac{1}{2}\ln(1+\sqrt{1+z^2}).
\label{f2surfsp}
\end{eqnarray}
For the leading correction terms $f_0^{\rm{span}}(z\xi)$ we obtain
\begin{eqnarray}
f_0^{\rm{span}}(z \xi)&=&\frac{1}{4}\ln{S}-\frac{1}{4}\ln{\xi}-\ln {\eta(i z\xi)}-\frac{1}{2}\ln{2}-\frac{1}{4}\ln{(1+z^2)}
\nonumber\\
&=&\frac{1}{4}\ln{S}-\frac{1}{2}\ln {\eta(i z\xi)\eta(i/(z\xi))}-\frac{1}{2}\ln{2}-\frac{1}{4}\ln{(1+z^2)}+\frac{1}{4}\ln{z},
\label{f0freesp}
\end{eqnarray}
in which  $\xi = \frac{N}{M}$.
Here we use the behavior of the Dedekind  eta function under the Jacobi transformation $\tau \to \tau'=-1/\tau$ for $\tau = i z \xi$ (see Eq. (\ref{jacobi})). For subleading correction terms $f_p^{\rm{span}}(z \xi)$ for $p=1, 2, 3,
..., $ we obtain
\begin{eqnarray}
f_p^{\rm{span}}(z \xi)&=&\frac{\pi^{2p+1} \xi^{p+1}}{2^{2p}}\frac{\Lambda_{2p}}{(2p)!}
\frac{ K_{2p+2}^{0,0}(i z \xi)}{2p+2}. \nonumber
\end{eqnarray}
As an example, we will list expansion coefficients $f_p^{\rm{span}}(z \xi)$
for $p=1,~2$
\begin{eqnarray}
f_1^{\rm{span}}(z\xi)&=& z(1+z^2)\frac{\pi^3 \xi^2 }{1080}\;(\theta
_2^4\theta _3^4+\theta _3^4\theta _4^4-\theta _2^4\theta _4^4),
\nonumber\\
f_2^{\rm{span}}(z\xi)&=&\frac{\pi^5 \xi^3
}{12096}\left[\frac{z(1+z^2)(1+9z^2)}{5}+\frac{z^2(1+z^2)^2}{3}\frac{\partial}{\partial z}\right]
(\theta_2^4+\theta_3^4)(\theta_2^4-\theta_4^4)(\theta_3^4+\theta_4^4), \label{fp12sp}
\end{eqnarray}
where $\eta$ is the Dedekind eta function and $\theta_i =
\theta_i(z \xi)$ with $i=2,3,4$.

From Eqs. (\ref{f0free}), (\ref{f0cyl}) and (\ref{f0freesp}) we can see that universal part of the $f_0$ is given  by Eq. (\ref{funiv}) with central charge $c=-2$.
This proves the conformal field theory prediction for  the corner free energy and shows that the corner free energy, which is proportional to the central charge $c$, is indeed universal.

It is interesting to note that leading finite-size corrections $f_0$ for the
dimer model on a $(2M-1) \times (2N-1)$ square lattice with single monomer on
the boundary is similar for both free and cylindrical boundary conditions. This similarity is unusual in integrable models and requires explanation \cite{izm2005}.
Consider the dimer model on a $(2M-1) \times (2N-1)$ square lattice ${\cal L}$ with a single monomer on the boundary and with periodic boundary condition in the horizontal direction and free boundary conditions in the vertical direction. The lattice ${\cal L}$ then forms a cylinder of perimeter $(2N-1)$ and height $(2M-1)$. Let us enumerate the sites of the lattice ${\cal L}$ as $(m,n)$, where $m=1, 2, ..., 2N-1$ and $m=1, 2, ..., 2M-1$. There is a bijection between dimer coverings on ${\cal L}$ with one boundary site  removed and spanning trees on the odd - odd sublattice $G
\subset {\cal L}$ with sites labeled as $(2n-1,2m-1)$ with $n=1, 2, ..., N$ and $m=1,2,...,M$.

Let us select the odd-odd sublattice $G$ and put monomer on the boundary.  It is easy to see that
two columns of $G$ will be neighbours in $G$ and in ${\cal L}$
(connected by horizontal bonds). Therefore a dimer may cover zero,
one or two sites of $G$. The dimers covering no site of
$G$ are completely fixed by the others and play no role. For the
others, we do the following construction. If
a dimer touches only one site of $G$, we draw an arrow directed
along the dimer from that site to the nearest neighbouring site of
$G$. However, for a dimer laid on two sites of $G$, the two arrows
would point from either site to the other, ruining the spanning
tree picture. It can nevertheless be restored in the following
way. Instead of seeing the two arrows as pointing from one site to its
neighbour, we say that they point towards roots inserted between
the neighbour sites, thus replacing the arrows \mbox{\arrowsite}
by \mbox{\arrowroot}. This in effect amounts to opening the
cylinder by removing the horizontal bonds of~${\cal L}$ which connect
sites of $G$, unwrapping it into a strip, and to adding columns of
roots on the left and on the right side of the strip. The new
arrow configurations define spanning trees, rooted anywhere on the
left and right boundaries. So dimer coverings on the original
cylinder are mapped to spanning trees on a strip, with close
horizontal boundaries, and open vertical boundaries.
Therefore, although the dimer model is
originally defined on a cylinder, it shows the finite-size
corrections expected on a strip, and must really be viewed as a
model on a strip.

For odd - even $(2M-1)\times 2N$ and even - even
$2M \times 2N$ cylinders with perimeter $2N$, the problem of having two arrows pointing from and to neighbor sites does not arise; however, the arrows one obtains do not define spanning trees but rather a combination of loops wrapped around the cylinder and tree branches attached to the loops.

\section{Conclusion}
\label{conclusion}

We have derived the exact finite-size corrections for the free energy of the anisotropic dimer model on the $(2M-1) \times (2N-1)$ rectangular lattice with free and cylindrical boundary conditions with a single monomer on the boundary and for the free energy of the anisotropic spanning tree on the $M \times N$ rectangular lattice. We found that the exact asymptotic expansion of the free energy of the dimer model and spanning tree can be written in the form given by Eq. (\ref{Fnew}). We also show that the dimer model on the cylinder with an odd number of sites on the perimeter exhibits the same finite-size corrections as on the plane.

We  proved the conformal field theory prediction about the corner free energy and have shown that the corner free energy, which is proportional to the central charge $c$, is indeed universal. We find the central charge in the framework of the conformal field theory to be $c=-2$.

\section{Acknowledgment}
\label{Acknowledgment}

W.G. wishes to thank J.L. Jacobsen for drawing our attention to this subject. The work of W.G. and X.W. were supported by the National Science Foundation of China under Grant No. 11175018. The work of N.I. and R.K. were supported by a Marie Curie IIF (Project no. 300206-RAVEN)and IRSES (Projects no. 295302-SPIDER and 612707-DIONICOS) within 7th European Community Framework Programme and by the grant of the Science Committee of the Ministry of Science and Education of the Republic of Armenia under contract 13-1C080.

\appendix

\section{Asymptotic expansion of $Z_{0,0}(z,M,N)$}
\label{Expansion Z(0,0)(z,M,N)}

In this section we shall obtain the exact asymptotic expansion of
the logarithm of the  $Z_{0, 0}(z,M,N)$. The logarithm of the  $Z_{0, 0}(z,M,N)$ can be expanded in the similar way as it has been done in Ref. \cite{izm2002} for the isotropic dimer model ($z_h = z_v$ or $z=1$).

$Z_{0,0}(z,M,N)$ can be transform in the following way
\begin{eqnarray}
Z^2_{0,0}(z,M,N)&=&
\prod_{n=0}^{N-1}{\prod_{m=0}^{M-1}}^{\prime}4\left[\textstyle{
\;z^2 \sin^2\frac{m \pi}{{M}}+ \sin^2\frac{n \pi}{N}}\right]
\nonumber\\
&=& \left(\prod_{n=1}^{N-1}4\;\textstyle{\sin^2\frac{n
\pi}{N}}\right)
\prod_{n=0}^{N-1}{\prod_{m=1}^{M-1}}4\left[\textstyle{ \;z^2
\sin^2\frac{m \pi}{M}+ \sin^2\frac{n \pi}{N}}\right]
\nonumber\\
&=& N^2 \prod_{n=0}^{N-1}{\prod_{m=1}^{M-1}}4\left[\textstyle{
\;z^2 \sin^2\frac{m \pi}{M}+ \sin^2\frac{n \pi}{N}}\right].
\label{Z001}
\end{eqnarray}
With the help of the identity given by Eq. (\ref{identity1}),
$Z_{0,0}(z,M,N)$ can be transformed into a simpler form
\begin{equation}
Z_{0,0}(z,M,N)=N\;\prod_{m=1}^{M-1} 2 \textstyle{~\! {\rm
sinh}\left[N\; \omega_z\!\left(\frac{m \pi}{M}\right)\right] }\,
\label{Zab1}
\end{equation}
where lattice dispersion relation is
\begin{equation}
\omega_{z}(k)={\rm arcsinh}\left(z \sin{k}\right).
\label{SpectralFunction}
\end{equation}

Considering the logarithm of $Z_{0,0}(z,M,N)$, we note, that it
can be transformed as
\begin{equation}
\ln Z_{0,0}(z,M,N)=\ln{N}+ N\sum_{m=1}^{M-1}
\omega_z\!\left(\textstyle{\frac{\pi m}{M}}\right)+
\sum_{m=1}^{M-1}\ln\,\big[1-e^{-2\,N \omega_z\left(\frac{\pi m
}{M}\right)\,}\big]. \label{lnZab}
\end{equation}
The second sum here vanishes in the formal limit $N\to\infty$ when
the system turns into infinitely long strip of width $M$. The
asymptotic expansion of the first sum can be found with the help
of the Euler-Maclaurin summation formula \cite{Abramowitz}
\begin{equation}
N\sum_{m=0}^{M-1}\omega_z\!\left(\textstyle{\frac{\pi
m}{M}}\right)= \frac{S}{\pi}\int_{0}^{\pi}\!\!\omega_z(x)~\!{\rm
d}x-\pi z \xi\,{\rm B}_{2}- 2\pi\xi\sum_{p=1}^{\infty}
\left(\frac{\pi^2\xi}{S}\right)^{p}
\frac{z_{2p}}{(2p)!}\;\frac{{\rm B}_{2p+2}}{2p+2},
\label{EulerMaclaurinTerm}
\end{equation}
where ${\rm B}_{p}$ are so-called Bernoulli numbers ($B_2=1/6,
B_4=-1/30, B_6=1/42, ...$) and
\begin{equation}
\int_{0}^{\pi}\!\!\omega_z(x)~\!{\rm d}x =
\frac{1}{2}\Phi(-z^2,2,\frac{1}{2}),
\end{equation}
 where $\Phi(x,s,\alpha)$ is
the Lerch transcendent defined as
\begin{equation}
\Phi(x,s,\alpha)=\sum_{n=0}^{\infty}(\alpha+n)^{-s}x^n.
\label{phi}
\end{equation}
In particular, for isotropic dimer model ($z=1$), the Lerch transcendent is now $\Phi(-1,2,{1}/{2})=4 G$, where $G$ is the Catalan constant given by
\begin{equation}
 G=\sum_{n=0}^{\infty}{
                      \frac{(-1)^n}{(2n+1)^2}
                      }=0.915 965 594 \dots.
\label{Catelan}
\end{equation}
We have also used the symmetry property,
$\omega_z(k)=\omega_z(\pi-k)$, of the lattice dispersion relation
(\ref{SpectralFunction}) and its Taylor expansion
\begin{equation}
\omega _z(k)=k\left(z +\sum _{p=1}^{\infty} \frac{z
_{2p}}{(2p)!}k^{2p}\right),
\label{SpectralFunctionExpansion}
\end{equation}
with $z_2=-z(1+z^2)/3$, $z_4=z(1+z^2)(1+9z^2)/5$,
$z_6=-z(1+z^2)(1+90z^2+225z^4)/7$, etc.

The second sum in Eq. (\ref{lnZab}) can be analyzed in the
following way: we first expand $\ln(1-e^A)$ as a power series in
$e^A$ and then  split the sum in two part: $m \in [0,[M/2]-1]$ and
$m \in [[M/2], M-1]$ and finally we change variable m in the
second part viz $m \to M - m$. As result we
obtain
\begin{eqnarray}
\sum_{m=1}^{M-1}\ln\big[\,1-e^{-2\,N\omega_z\left(\frac{\pi m
}{M}\right)}\big]&&=-\sum_{n=1}^{\infty}\frac{1}{n}\left\{\sum_{m=1}^{[M/2]-1}
e^{-2n\,N\omega_z\left(\frac{\pi m}{M}\right)
\big]}+\sum_{m=1}^{M-[M/2]} e^{-2n\,N\omega_z\left(\frac{\pi
 m}{M}\right)\big]}\right\}. \label{Z2exp}
\end{eqnarray}
The argument of the  exponent can be expanded in powers of $1/S$
if we replace the lattice dispersion relation $\omega_z(x)$ with
its Taylor expansion (\ref{SpectralFunctionExpansion})
$$ \exp\left[-2 n N \omega_z\left(\frac{\pi m}{M}\right)\right]=\exp\left\{-2\pi
m n z \xi-2\pi n \xi \sum_{p=1}^{\infty}\frac{z_{2p}}{(2p)!}
\left(\frac{\pi^2\xi}{S}\right)^{p}m ^{2p+1}\right\},$$
where $\xi=M/N$. Taking into account the relation between moments
and cumulants (see Appendix \ref{MomentsCumulants}), we obtain
asymptotic expansion of the first exponent itself in powers of
$1/S$
\begin{eqnarray}
e^{-2 n N \omega_z\left(\frac{\pi m}{M}\right)}= e^{-2\pi n m z
\xi} -2\pi n \xi\sum_{p=1}^{\infty}
\left(\frac{\pi^2\xi}{S}\right)^{p}\frac{\Lambda_{2p}}{(2p)!}
~m^{2p+1}e^{-2\pi n m z \xi}. \nonumber
\end{eqnarray}
The differential operators $\Lambda_{2p}$ that have appeared here
can be expressed via coefficients $z_{2p}$ of the expansion of the
lattice dispersion relation as
\begin{eqnarray}
{\Lambda}_{2}&=&z_2,\nonumber\\
{\Lambda}_{4}&=&z_4+3z_2^2\,\frac{\partial}{\partial z},\nonumber\\
{\Lambda}_{6}&=&z_6+15z_4 z_2\,\frac{\partial}{\partial }
+15z_2^3\,\frac{\partial^2}{\partial z^2}. \nonumber\\
&\vdots&\nonumber
\end{eqnarray}
Plugging the expansion of the exponent back into Eq. (\ref{Z2exp})
we obtain
\begin{eqnarray}
&&\sum_{m=1}^{M-1}\ln\left[\,1-e^{-2 N\omega_z\left(\frac{\pi
m}{M}\right)}\right]=-\;\sum_{n=1}^{\infty}\frac{1}{n}\left\{
\sum_{m=1}^{[M/2]-1}e^{-2\pi n m z \xi }+\!\!\!
\sum_{m=1}^{M-[M/2]}\!\!\!e^{-2\pi n m z \xi}\right\}\nonumber\\
&&~+~2\pi\xi\sum_{p=1}^{\infty}
\left(\frac{\pi^2\xi}{S}\right)^{p}\frac{\Lambda_{2p}}{(2p)!}\;
\sum_{n=1}^{\infty}\left\{\sum_{m=1}^{[M/2]-1} m^{2p+1}~e^{-2\pi n
m z \xi}+\sum_{m=1}^{M-[M/2]} m^{2p+1}~e^{-2\pi n m z \xi}
\right\}.\nonumber
\end{eqnarray}
In all these series, summation over $m$ can be extended to
infinity. The resulting errors are exponentially small and do not
affect our asymptotic expansion in any finite power of $1/S$. As
result we obtain
\begin{eqnarray}
\sum_{m=1}^{M-1}\ln\left[\,1-e^{-2 N\omega_z\left(\frac{\pi
m}{M}\right)}\right]&=&-2\;\sum_{n=1}^{\infty}
\sum_{m=1}^{\infty}\frac{1}{n}e^{-2\pi n m z \xi}\nonumber\\
&&~+~4\pi\xi\sum_{p=1}^{\infty}
\left(\frac{\pi^2\xi}{S}\right)^{p}\frac{\Lambda_{2p}}{(2p)!}\;
\sum_{n=1}^{\infty} \sum_{m=1}^{\infty} m^{2p+1}~e^{-2\pi n m z
\xi}.
\nonumber
\end{eqnarray}
The key point of our analysis is the observation that all the
series that have appeared in such an expansion can be obtained by
resummation of either the Dedekind eta  function, $\eta(\tau)$, or
Kronecker's double series, ${\rm K}_{p}^{0,0}(\tau)$.

The Dedekind eta function is usually defined as
$$\eta(\tau)=e^{\pi i\tau/12}\prod_{n=1}^{\infty}\Big[\,1-e^{2\pi
i \tau n}\,\Big].$$
Considering the logarithm of $\eta(\tau)$ of
pure imaginary aspect ratio, $\tau=i\xi$, we obtain the identity
\begin{eqnarray}
\ln\eta(i\xi)+\frac{\pi\xi}{12}=-\sum_{n=1}^{\infty}\sum_{m=1}^{\infty}\frac{1}{m}e^{-2\pi m n \xi}.
\label{IdentityTheta}
\end{eqnarray}
Kronecker's double series can be defined as \cite{Weil}
$${\rm K}_{p}^{0,0}(\tau)= -\frac{p!}{(-2\pi i)^p}
{\sum_{m,n\in Z}}^{\prime} \frac{1}{(n+\tau m)^{p}}, $$
where the prime over product denotes the restriction $(m,n) \ne
(0,0)$. In this form, however, they cannot be directly applied to
our analysis. We need to cast them in a different form. The final
result of our resummation of double Kronecker sum is
$${\rm K}_{p}^{0,0}(\tau)={\rm B}_{p}-p\sum_{m\neq
0}\sum_{n=0}^{\infty} n^{p-1}~e^{2\pi i m n \tau}.$$
Considering the Kronecker sums with pure imaginary aspect ratio,
$\tau=i\xi$, we can further rearrange this expression to get
summation only over positive $m\geq 1$
\begin{eqnarray}
{\rm B}_{2p}-{\rm K}_{2p}^{0,0}(i\xi) &=&4p\;\sum_{m=1}^{\infty}
\sum_{n=1}^{\infty} n^{2p-1}~e^{-2\pi m n \xi}.
\label{IdentityKronecker}
\end{eqnarray}
Note, that Kronecker functions ${\rm K}_{2p}^{0,0}(\tau)$ can all
be expressed in terms of the elliptic $\theta$-functions only (see
Appendix \ref{KroneckerToTheta}).

Now, with the help of the identities (\ref{IdentityTheta}) and
(\ref{IdentityKronecker}) we obtain
\begin{eqnarray}
&&\sum_{m=0}^{M-1}\ln\left[\,1-e^{-2 N \omega_z\left(\frac{\pi m
}{M}\right)}\right]= 2 \ln \eta(i z\xi)+\pi z \xi\,{\rm
B}_2\nonumber\\[0.1cm]&&~~~~~~~~~~~~~~~~~-2\pi\xi\sum_{p=1}^{\infty}
\left(\frac{\pi^2\xi}{S}\right)^{p}\frac{\Lambda_{2p}}{(2p)!}
\,\frac{\;{\rm K}_{2p+2}^{0,0}(i z \xi)-{\rm B}_{2p+2}}{2p+2}.
\label{ln(1-e)}
\end{eqnarray}
Substituting Eqs. (\ref{EulerMaclaurinTerm}) and (\ref{ln(1-e)})
into Eq. (\ref{lnZab}) we finally obtain exact asymptotic
expansion of the logarithm of $Z_{0,0}(z,M,N)$ in terms of the
Kronecker's double series

\begin{eqnarray}
\ln Z_{0,0}(z,M,N)&=&\frac{S}{\pi}\int_0^{\pi} \omega_{z}(x)dx +
\ln\sqrt{S \xi}+ 2 \ln \eta(i z \xi)\nonumber\\
&-&2\pi\xi\sum_{p=1}^\infty \left(\frac{\pi^2
\xi}{S}\right)^p\frac{\Lambda_{2p}}{(2p)!} \frac{
K_{2p+2}^{0,0}(i z \xi)}{2p+2}, \label{ExpansZab}
\end{eqnarray}
where $S = M N$, $\xi = N/M$. Note, that Bernoulli numbers ${\rm
B}_{p}$ have finally dropped out from the asymptotic expansion.

\section{Relation between moments and cumulants}
\label{MomentsCumulants}
In this Appendix we consider the relation between moments $Z_k$ and cumulants $F_k$ which enters the expansion of exponent
$$\exp\left\{\;\sum_{k=1}^{\infty}\frac{x^k}{k!}\,F_k\,\right\}
=1+\sum_{k=1}^{\infty}\frac{x^k}{k!}\,Z_k,$$
These are related in the following manner \cite{Prohorov}
\begin{eqnarray}
Z_1&=&F_1,\nonumber\\
Z_2&=&F_2+F^2_1,\nonumber\\
Z_3&=&F_3+3F_1F_2+F^3_1,\nonumber\\
Z_4&=&F_4+4F_1F_3+3F_2^2+6F_1^2F_2+F^4_1,\nonumber\\
~&\vdots&~\nonumber\\ Z_k&=&\sum_{r=1}^{k},\sum
\left(\frac{F_{k_1}}{k_1!}\right)^{i_1},\ldots
\left(\frac{F_{k_r}}{k_r!}\right)^{i_r} \frac{k!}{i_1!\ldots
i_r!},\nonumber
\end{eqnarray}
where summation is over all positive numbers $\{i_1\ldots i_r\}$
and different positive numbers $\{k_1,\ldots,k_r\}$ such that $k_1
i_1+\ldots+ k_r i_r=k$.

\section{Reduction of Kronecker's double series to theta functions}
\label{KroneckerToTheta}

The Laurent expansion of the Weierstrass function is given by
\begin{eqnarray}
\wp(z)&=&\frac{1}{z^2}+\sum_{(n,m)\neq(0,0)}\left[\frac{1}{(z-n-\tau
m)^2}-\frac{1}{(n+\tau m)^2}\right]\nonumber\\
&=&\frac{1}{z^2}+\sum_{p=2}^{\infty}a_{p}(\tau) z^{2p-2}.\nonumber
\end{eqnarray}
The coefficients $a_p(\tau)$ of the expansion can all be written
in terms of the elliptic $\theta$-functions with the help of the
recursion relation \cite{Korn}
$$a_p=\frac{3}{(p-3)(2p+1)}~
(a_{2}a_{p-2}+a_{3}a_{p-3}+\ldots+a_{p-2}a_{2}),$$
where first terms of the sequence are
\begin{eqnarray}
a_2&=&\textstyle{\frac{\pi^4}{15}}(\theta_2^4\theta_3^4-
\theta_2^4\theta_4^4+\theta_3^4\theta_4^4),\nonumber\\
a_3&=&\textstyle{\frac{\pi^6}{189}}(\theta_2^4+\theta_3^4)
(\theta_4^4-\theta_2^4)(\theta_3^4+\theta_4^4),\nonumber\\
a_4&=&\textstyle{\frac{1}{3}}a_2^2,\nonumber\\
a_5&=&\textstyle{\frac{3}{11}}(a_2a_3),\nonumber\\
a_6&=&\textstyle{\frac{1}{39}}(2a_2^3+3a_3^2),\nonumber\\
&\vdots& .\nonumber
\end{eqnarray}
Kronecker functions ${\rm K}_{2p}^{0,0}(\tau)$ are related
directly to the coefficients $a_p(\tau)$
$${\rm K}^{0,0}_{2p}(\tau)=
-\frac{(2p)!}{(-4\pi^2)^p}\frac{a_p(\tau)}{(2p-1)}.$$

From the general formulas above we can easily write down all the
Kronecker functions that have appeared in our asymptotic
expansions in terms of the elliptic $\theta$ - functions, e.g.
\begin{eqnarray}
{\rm K}_{4}^{0,0}(\tau)&=&
\textstyle{\frac{1}{30}(\theta_2^4\theta_4^4-\theta_2^4\theta_3^4-\theta_3^4\theta_4^4)},\nonumber\\
 {\rm K}_{6}^{0,0}(\tau)&=&\textstyle{\frac{1}{84}
(\theta_2^4+\theta_3^4)(\theta_4^4-\theta_2^4)(\theta_3^4+\theta_4^4)},\nonumber\\
K_8^{0,0}(\tau)&=&\textstyle{-\frac{1}{30}\left[\theta_2^4\theta_3^4
-\theta_2^4\theta_4^4+\theta_3^4\theta_4^4\right]^2},\nonumber\\
K_{10}^{0,0}(\tau )&=&\textstyle{\frac{5}{132}
\left[\theta_2^4+\theta_3^4\right]\left[\theta_4^4+\theta_2^4\right]
\left[\theta_3^4+\theta_4^4\right]
\left[\theta_2^4\theta_3^4-\theta_2^4\theta_4^4+
\theta_3^4\theta_4^4\right]},\nonumber\\
&\vdots&.\nonumber
\end{eqnarray}
Note that when $\xi\to\infty$ we have limits $\theta_2\to 0$,
$\theta_4\to 1$, $\theta_3\to 1$ and the Kronecker's function
reduce to the Bernoulli polynomials.



\end{document}